\documentclass[aps,prc,letterpaper,11pt,twoside,tightenlines,nofootinbib,showpacs,preprint]{revtex4}
\usepackage[latin1]{inputenc}
\usepackage{amsmath}
\usepackage{amsfonts}
\usepackage{amssymb}
\usepackage{graphicx}
\usepackage{subfigure}
\usepackage{rotating}
\begin{document}
\title{Strategies for an accurate determination of the X(3872) energy from QCD lattice simulations}

\author{E. J. Garzon$^1$, R. Molina$^{2}$, A. Hosaka$^{2,3}$ and E. Oset$^1$ }
\affiliation{$^1$ Departamento de F\'{\i}sica Te\'orica and IFIC, Centro Mixto Universidad de Valencia-CSIC,
Institutos de Investigaci\'on de Paterna, Aptdo. 22085, 46071 Valencia,Spain}
\affiliation{$^2$ Research Center for Nuclear Physics (RCNP), Osaka University, Ibaraki, Osaka 567-0047, Japan}
\affiliation{$^3$ J-PARC Branch, KEK Theory Center, Institute of Particle and Nuclear Studies, KEK, Tokai, Ibaraki, 319-1106, Japan}

\date{\today}

\begin{abstract} 
We develop a method to determine accurately the binding energy of the X(3872) from lattice data for the $D \bar D^*$ interaction. We show that, because of the small difference between the neutral and charged components of the X(3872), it is necessary to differentiate them in the energy levels of the lattice spectrum if one wishes to have a precise determination of the the binding energy of the X(3872). The analysis of the data requires the use of coupled channels. Depending on the number of levels available and the size of the box we determine the precision needed in the lattice energies to finally obtain a desired accuracy in the binding energy.
\end{abstract}


\maketitle

\section{Introduction}

The X(3872) state, observed for the first time by Belle~\cite{Choi:2003ue}, has been found in many other experiments and is the paradigm of the charmonium states of non-conventional nature (see Refs.~\cite{Brambilla:2010cs,Voloshin:2007dx} for recent reviews on the issue). Although for some time the quantum numbers were not well determined and both, the $J^{PC}=1^{++}$ and $2^{-+}$ were candidates, theoretical papers showed a preference for the $1^{++}$ state~\cite{Swanson:2003tb,Tornqvist:2004qy,Braaten:2004rw,Gamermann:2007fi,Hanhart:2011tn}, which has been recently confirmed by the LHCb~\cite{Aaij:2013zoa}.

 The search for this state in lattice QCD simulations has also run parallel and several works have been devoted to this task~\cite{Liu:2012ze,Bali:2012ua,Bali:2011dc,Mohler:2012na}, finding one state close to the experimental one. Yet, it was too difficult to unambiguously determine whether one had a bound state or simply $D \bar D^*$ scattering states which appear at around the same energy. An important step has been given very recently in Ref.~\cite{Prelovsek:2013cra}, where a bound state is obtained in a dynamical $N_f=2$ lattice simulation with $11 \pm 7$ MeV below the $D \bar D^*$ threshold and quantum numbers $1^{++}$. Improvements on this can be done in the future using larger boxes and smaller pion masses. 
 
  The purpose of the present paper is to find a strategy to determine accurately the binding energy of the X(3872) in lattice QCD simulations. A precise determination, with an energy about 0.2 MeV below the $D^0 \bar D^{*0}$ threshold, requires to differentiate between the $u$ and $d$ quark masses in order to account for the 7 MeV difference between the neutral and charged components of the wave function~\cite{Gamermann:2009fv,Gamermann:2009uq}. The small binding of the state with respect to the $D^0 \bar D^{*0}$ threshold, much smaller than the difference of masses between the $D^0 \bar D^{*0}$ and $D^+ D^{*-}$ components, makes this consideration imperative in order to get a precise value of the binding energy and unambiguously determine the bound state character of the X(3872). In fact, when this is done, energy levels can be associated to either $D^0 \bar D^{*0}$ or $D^+ \bar D^{*-}$. 
  
  The strategy used here follows closely the work of Ref.~\cite{Doring:2011vk} using coupled channels, where the energy levels related to the scalar mesons were investigated. It studies the levels of two-meson interaction in a finite box and tackles the inverse problem of deriving phase shifts from pseudolattice data using L\"uscher formalism~\cite{Luscher:1990ux} and different strategies. The case of bound states is studied along similar lines in Ref.~\cite{Albaladejo:2013aka}, where the combination of L\"uscher formalism and methods related to those used in Refs.~\cite{Beane:2003da,Beane:2011iw} allow a precise determination of binding energies of hidden charm states. Thus, in order to have an accurate measurement of the binding energy of the X(3872), we present a method using different number of levels, different box sizes and determine the precision required for the lattice energies.

\section{The X(3872) in the continuum limit}

\begin{figure}
\includegraphics[scale=0.5]{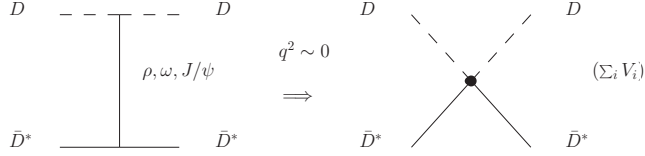}
\caption{Point-like pseudoscalar - vector interaction.}
\label{fig:PV1}
\end{figure}

In this section we discuss briefly the dynamical generation of the X(3872) in the continuum limit. All the details are in Refs.~\cite{Gamermann:2007fi, Gamermann:2009fv,Aceti:2012cb}. The pseudoscalar - vector interaction can be studied through the hidden gauge Lagrangian~\cite{bando}, which contains interaction between vectors and with pseudoscalar mesons,
\begin{equation}
{\cal L}_{III}=-\frac{1}{4}\langle V_{\mu \nu}V^{\mu\nu}\rangle+
\frac{1}{2}M^2_V \langle [ V_\mu-\frac{i}{g} \Gamma_\mu ] \rangle
\end{equation}
where $V_{\mu\nu}=\partial_{\mu} V_\nu -\partial_\nu V_\mu -ig[V_\mu,V_\nu]$, and
$g=\frac{M_V}{2f}$. The model is based on vector-meson exchange, see Fig.~\ref{fig:PV1}. From the above equation, the lower and upper vertices needed to evaluate the amplitude of the diagram depicted in Fig.~\ref{fig:PV1} are obtained using the terms
\begin{eqnarray}
\mathcal{L}_{PPV}=-ig\langle  V^\mu [P,\partial_\mu P]\rangle,\hspace{1cm}
\mathcal{L}_{3V} = ig\langle (V^\mu \partial_\nu V_\mu -\partial_\nu V_\mu V^\mu)
V^\nu\rangle\ ,\label{lag3}
\end{eqnarray}
where $V_\mu$, $P$ are the matrices of the 16-plet of vector, pseudoscalar mesons~\cite{Gamermann:2009fv}. In fact, the combination of both terms in Eq. (\ref{lag3}), for s-wave, when the momenta $q^2$ exchanged in the propagator of the vector meson exchanged can be neglected against $-M_V^2$, leads to a point-like interaction, and is equivalent to using the Lagrangian,
\begin{equation}
{\cal L}_{PPVV}=-{\frac{1}{4f^2}}Tr\left(J_\mu\cal{J}^\mu\right). 
\label{lag}
\end{equation}
with $J_\mu=(\partial_\mu P)P-P\partial_\mu P $ and $\cal{J}_\mu=(\partial_\mu \cal{V}_\nu)\cal{V}^\nu-\cal{V}_\nu\partial_\mu \cal{V}^\nu$, 
see~\cite{Roca:2005nm,Gamermann:2009fv}. In Ref.~\cite{Gamermann:2009fv}, the currents in Eq. (\ref{lag}) are separated for heavy and light vector-meson-exchange, introducing the breaking parameters,
\begin{eqnarray}
\gamma=\left(\frac{m_{8^*}}{m_{3^*}}\right)^2=\frac{m^2_L}{m^2_H}\nonumber\hspace{1cm}
\psi=  \left(\frac{m_{8^*}}{m_{1^*}}\right)^2=\frac{m^2_L}{m^2_{J/\psi}}\label{break}\ ,
\end{eqnarray}
with $m_{8^*}=m_L=800$ MeV, $m_{3^*}=m_H=2050$ MeV and $m_{1^*}=m_{J/\psi}=3097$ MeV. This gives, $\gamma=0.14$ and $\psi=0.07$. Because of the smallness of the breaking parameters, the light and heavy sector are almost disconnected, and the transition potential between those is very small. Also, for light mesons, $f=f_\pi=93$ MeV, and for heavy ones, $f=f_D=165$ MeV, is used. Thus, the amplitude of the process $V_1(k)P_1(p)\to V_2(k\rq{}) P_2(p\rq{})$, is given by
\begin{equation}
V_{ij}(s,t,u)=\frac{\xi_{ij}}{4f_i f_j}(s-u)\,\vec{\epsilon} . \vec{\epsilon} ' \label{ampli}
\end{equation}
with $s-u=(k+k\rq{})(p+p\rq{})$, which must be projected in s-wave \cite{Gamermann:2009fv,Roca:2005nm}, and $i,j$ refer to the particle channels. Working in the charge basis, we have the channels $\frac{1}{\sqrt{2}}(\bar{K}^{*-}K^+-c. c.)$, $\frac{1}{\sqrt{2}}(\bar{K}^{*0}K^0-c. c.)$, $\frac{1}{\sqrt{2}}(D^{*+}D^--c. c.)$, $\frac{1}{\sqrt{2}}(D^{*0}\bar{D}^0-c. c.)$ and $\frac{1}{\sqrt{2}}(D^{*+}_s D^-_s-c. c.)$, and the matrix $\xi$ can be written in this basis as
\begin{equation}
\xi=\left(\begin{array}{ccccc}
-3&-3&0&-\gamma&\gamma\\-3&-3&-\gamma&0&\gamma\\0&-\gamma&-(1+\psi)&-1&-1\\
-\gamma&0&-1&-(1+\psi)&-1\\\gamma&\gamma&-1&-1&-(1+\psi)\\\end{array}\right)\ .
\end{equation}
Eq. (\ref{ampli}) is the input of the Bethe Salpether equation,
\begin{equation}
T={(I-VG)}^{-1} V\,\vec{\epsilon} . \vec{\epsilon} ' \ .\label{eq:bs}
\end{equation}
Here $G$ a diagonal matrix of the two-meson loop function for each channel. Usually it is evaluated with dimensional regularization and depends on the parameter $\alpha$~\cite{Gamermann:2009fv} ($\mu$ is a scale mass, fixed a priori) 
\begin{eqnarray}
G=G^{DR}(\sqrt{s}) &=& \frac{1}{16 \pi^2} \left\{ \alpha(\mu) + \ln
\frac{m_1^2}{\mu^2} + \frac{m_2^2-m_1^2 + s}{2s} \ln \frac{m_2^2}{m_1^2} +
\right. \nonumber \\ 
& &  +
\frac{q}{\sqrt{s}}
\left[
 \ln( s-(m_2^2-m_1^2)+2 q\sqrt{s})+
 \ln( s+(m_2^2-m_1^2)+2 q\sqrt{s}) \right. \nonumber  \\
& & \left. \left. 
-\ln(-s+(m_2^2-m_1^2)+2 q\sqrt{s})
-\ln(-s-(m_2^2-m_1^2)+2 q\sqrt{s}) \right]
\right\} \ ,
\label{eq:gpropdr}
\end{eqnarray}
One can also evaluate the $G$ function with a cuttoff, 
\begin{equation}
G=G^{co}(P_0=\sqrt{s}) =\int_{q <q_{max}} \frac{d^3 q}{(2\pi)^3} \frac{\omega_1 + \omega_2}{2 \omega_1  \omega_2}\frac{1}{(P^0)^2-(\omega_1 + \omega_2)^2+i\epsilon}
\label{eq:gfree}
\end{equation}

The calculation in Ref.~\cite{Gamermann:2009fv} is redone to get a binding energy more realistic at $0.2$ MeV with respect to the channel $D^{*0}\bar{D}^0-c. c.$~\cite{Aceti:2012cb}, where the masses of the mesons are taken from the PDG~\cite{pdg}. The free parameter, $\alpha$, is fixed for the light channels, $\alpha_L=-0.8$~\cite{Gamermann:2009fv, Roca:2005nm}, but the pole position of the X(3872) is not sensitive to that, since its mass is far away from these thresholds. For the heavy channels, the value $\alpha_H=-1.265$ is needed for such binding energy ($\mu$ is taken equal to 1500 MeV in all channels). In Table~\ref{tab:x3872}, a summary of the pole position and couplings of the resonance to each channel is given.
\begin{table}[htpb]
\centering
\begin{tabular}{lr}\hline\hline
\multicolumn{2}{l}{$\sqrt{s}_0=(3871.6-i 0.001)$ MeV}  \\
\hline
Channel&$|g_i|$ [MeV] \\
\hline
$\frac{1}{\sqrt{2}}(K^{*-}K^{+}-c.c)$&$53$\\
$\frac{1}{\sqrt{2}}(\bar{K}^{*0}K^0-c.c)$&$49$\\
$\frac{1}{\sqrt{2}}(D^{*+}D^{-}-c.c)$&$3638$\\
$\frac{1}{\sqrt{2}}(D^{*0}\bar{D}^0-c.c)$&$3663$\\
$\frac{1}{\sqrt{2}}(D^{*+}_{s} D^{-}_{s}-c.c)$&$3395$\\
\hline
\end{tabular}
\caption{Couplings of the pole  at $\sqrt{s}_0$ MeV to the channel $i$.}
\label{tab:x3872}
\end{table}

The Weinberg compositeness condition~\cite{Wein2} can be generalized for dynamically generated resonances from several channels~\cite{Gamermann:2009uq}, 
\begin{equation}
-\sum_i g^2_i \frac{\partial G}{\partial s}=1\ ,\label{wein2}
\end{equation}
being $s=P^2_0$, the squared of the initial energy in the center-of-mass frame, and $|g_i|$, the couplings in Table~\ref{tab:x3872}. Each term in Eq. (\ref{wein2}) gives the probability of finding
the $i$ channel in the wave function, which are $0.86$ for $D^{*0}\bar{D}^0-c.c$, 0.124 for $D^{*+}D^--c.c$ and $0.016$ for $D^{*+}_sD^-_s-c.c$.
However, this is different from the wave function at the origin $(2\pi)^{3/2} \psi (0)_i=g_i G_i$, which usually enters the evaluation of observables and are nearly equal \cite{Gamermann:2009uq}.

\section{Formalism in finite volume}
We follow the formalism used Ref.~\cite{Doring:2011vk} where the infinite volume amplitude $T$ is replaced by the amplitude $\tilde{T}$ in a finite box of size $L$ and $G(P^0)$ in Eqs. (\ref{eq:gpropdr}) and (\ref{eq:gfree}) is replaced by the finite volume loop function denoted with $\tilde{G}$, given by the discrete sum over eigenstates of the box
\begin{equation}
\tilde{G}(P^0)=\frac{1}{L^3} \sum_{\vec{q_i}} I(P^0, \vec{q}_i)
\label{eq:gbox}
\end{equation}
with
\begin{equation}
I(P^0, \vec{q}_i)= \frac{\omega_1(\vec{q_i}) + \omega_2(\vec{q_i})}{2 \omega_1(\vec{q_i})  \omega_2(\vec{q_i})}\frac{1}{(P^0)^2-(\omega_1(\vec{q_i}) + \omega_2(\vec{q_i}))^2}
\label{eq:iq}
\end{equation}
where $\omega_i=\sqrt{m_i^2+|\vec{q_i}~|^2}$ is the energy and the momentum $\vec{q}$ is quantized as
\begin{equation}
\vec{q}_i = \frac{2\pi}{L} \vec{n}_i
\end{equation} 
corresponding to the periodic boundary conditions. Here the vector $\vec{n}$, denotes the three dimension vector of all integers ($\mathbb{Z}^3$). 
This form produces a degeneracy for the set of three integer which has the same modulus. And we can write the modulus of the momentum as 
\begin{equation}
|\vec{q}_i| = \frac{2\pi}{L} \sqrt{m_i}
\label{eq:qn}
\end{equation}
where $m_i$ stands for the natural numbers ($\mathbb{N}$), and the multiplicity of the degeneracy is conveniently introduced in Eq. (\ref{eq:gbox}).
The sum over the momenta is done until a $q_{max}$, so the three dimension sum over $\vec{n}_i$ in Eq.~(\ref{eq:gbox}) becomes a one dimension sum over $m_i$ to an $n_{max}$ in a symmetric box
\begin{equation}
n_{max}=\frac{q_{max} L}{2 \pi}
\label{eq:qmax}
\end{equation}
When the dimensional regularization is used in the infinite volume case, as in section II, there is no trace of $q_{max}$ ($\alpha$ is related to $q_{max}$). Thus the equivalent formalism in finite volume should also be made independent of $q_{max}$ and related to $\alpha$. This is done in Ref.~\cite{MartinezTorres:2011pr} with the result
\begin{equation}
\tilde{G}=G^{DR}+\lim_{q_{max} \rightarrow \infty} \left( \frac{1}{L^3}\sum_{q<q_{max}} I(P^0,\vec{q}) - \int_{q<q_{max}} \frac{d^3 q}{(2\pi^3)} I(P^0,\vec{q}) \right) \equiv G^{DR} + \lim_{q_{max} \rightarrow \infty} \delta G
\label{eq:gtil}
\end{equation}
where $\delta G \equiv \tilde{G}-G^{co}$, and $G^{co}$ is given explicitly by the formula of Eq. (\ref{eq:gco}) \citep{Oller:1998hw}. Here $I(P^0,\vec{q})$ is the factor given in Eq. (\ref{eq:iq})
\begin{eqnarray}
G^{co}&=&\frac{1}{32 \pi^2} 
\left[ -\frac{\Delta}{s}log\frac{M_1^2}{M_2^2} +\frac{\nu}{s}
\left\lbrace
 log\frac{s-\Delta+\nu\sqrt{1+\frac{M_1^2}{q_{max}^2}}}{-s+\Delta+\nu\sqrt{1+\frac{M_1^2}{q_{max}^2}}} \right. \right. \nonumber \\
& &\left. +log\frac{s+\Delta+\nu\sqrt{1+\frac{M_2^2}{q_{max}^2}}}{-s-\Delta+\nu\sqrt{1+\frac{M_2^2}{q_{max}^2}}} +2\frac{\Delta}{s}log\frac{1+\sqrt{1+\frac{M_1^2}{q_{max}^2}}}{1+\sqrt{1+\frac{M_2^2}{q_{max}^2}}} \right\rbrace \nonumber \\ 
& &\left. -2log\left[\left(1+\sqrt{1+\frac{M_1^2}{q_{max}^2}} \right) \left(1+\sqrt{1+\frac{M_1^2}{q_{max}^2}} \right)\right] +log\frac{M_1^2 M_2^2}{q_{max}^4} \right]
\label{eq:gco}
\end{eqnarray}
where $\Delta=M_2^2-M_1^2$ and $\nu=\sqrt{\left[s-(M_1+M_2)^2 \right]\left[s-(M_1-M_2)^2 \right]}$.

\begin{figure}
\includegraphics[scale=0.7]{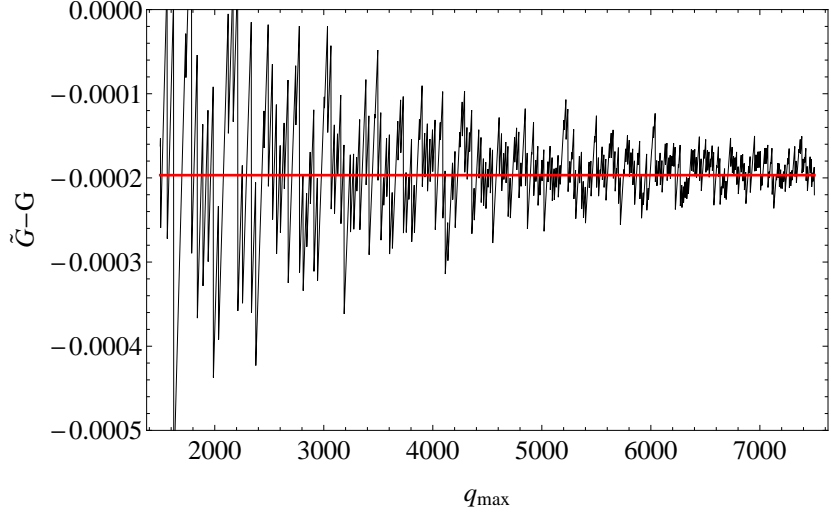}
\caption{Representation of $\delta G=\tilde{G}-G$ for $D^{+}D^{*-}$ in function of $q_{max}$ for $\sqrt{s}=3850$ MeV. The thick line represents the average of $\delta G$ for different values of $q_{max}$ between 1500 and 2500 MeV.}
\label{fig:deltag}
\end{figure}

\begin{figure}
\subfigure[$D^{0}\bar{D}^{*0}$]{\includegraphics[scale=0.5]{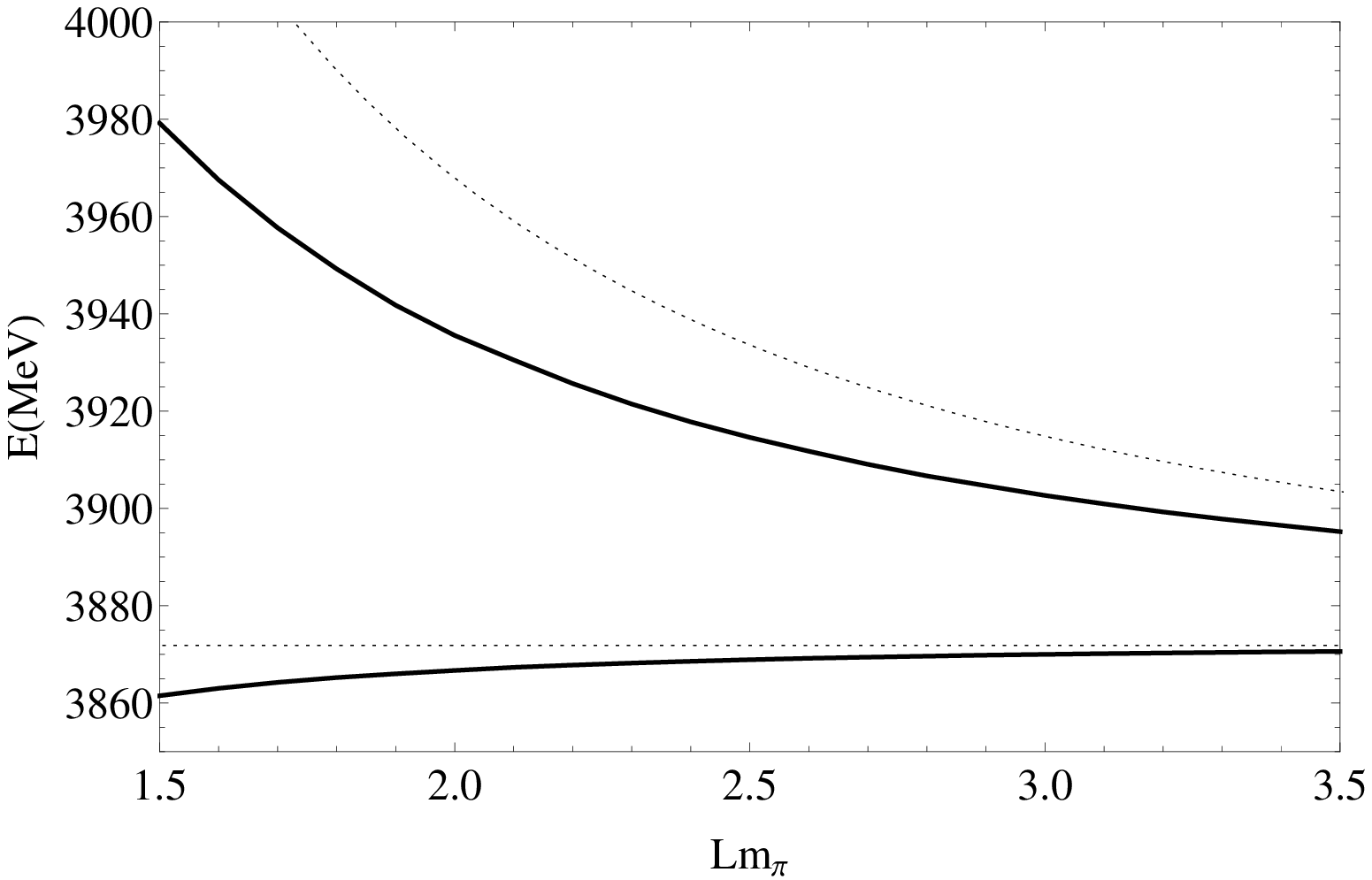}}
\subfigure[$D^{+}D^{*-}$]{\includegraphics[scale=0.5]{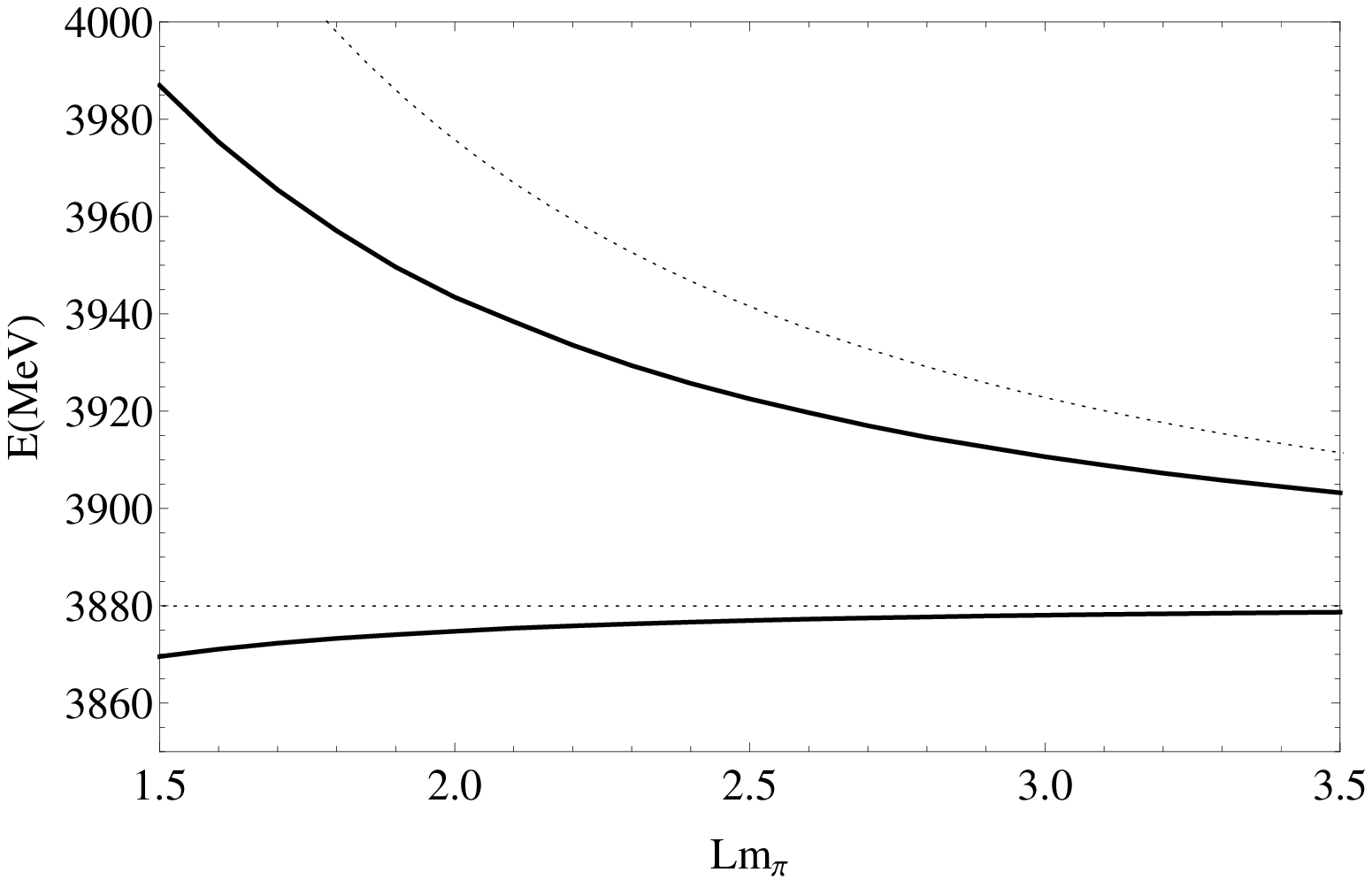}}
\caption{$L$ dependence of the energies of the poles for the two first levels of a single channel. Dotted lines correspond to the free energies.}
\label{fig:leddst}
\end{figure}

In Fig.~\ref{fig:deltag} we show that $\delta G$ converges as $q_{max} \rightarrow \infty$. In practice, one can take an average for different values between $q_{max}=$1500-2500 MeV and one sees that it reproduces fairly well the limit of $q_{max} \rightarrow \infty$.
In the present case we use $f_D=160$ MeV in the potential $V$.
The Bethe-Salpeter equation in finite volume, can be written as,
\begin{equation}
\tilde{T}=(I-V \tilde{G})^{-1}V
\end{equation}
or
\begin{equation}
\tilde{T}^{-1}=V^{-1}-\tilde{G}
\label{eq:bsbox}
\end{equation}
\begin{figure}
\includegraphics[scale=1.0]{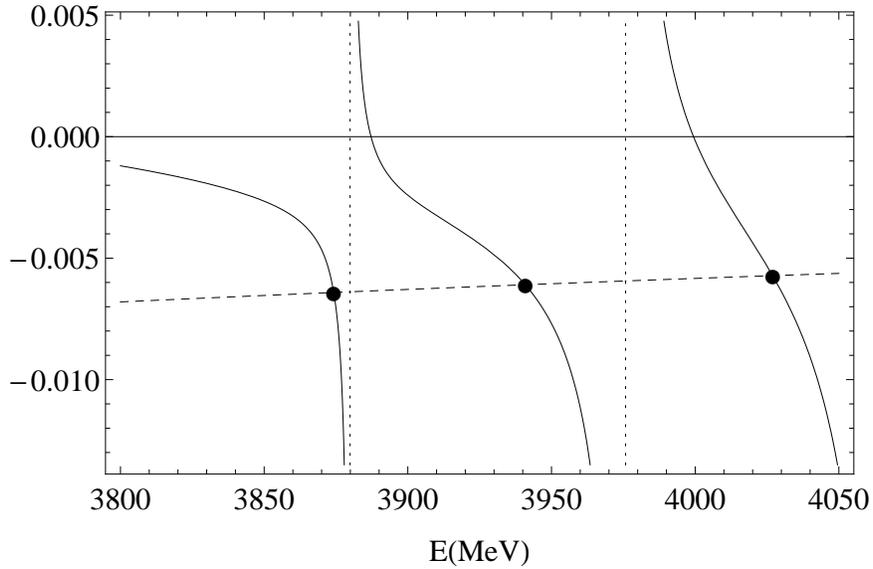}
\caption{$\tilde{G}$(solid) and $V^{-1}$(dashed) energy dependence of $D^{+}D^{*-}$ for $Lm_\pi=2.0$. Black dots correspond to energies ($E\equiv P^0$) where $V^{-1}=\tilde{G}$. Vertical dotted lines are the free energies in the box for $DD^*$.}
\label{fig:vg}
\end{figure}
The energy levels in the box in the presence of interaction $V$ correspond to the condition
\begin{equation}
det(I-V\tilde{G})=0.
\label{eq:tpole}
\end{equation}
In a single channel, Eq. (\ref{eq:tpole}) leads to poles in the $\tilde{T}$ amplitude when  $V^{-1}=\tilde{G}$. In Fig. (\ref{fig:vg}) we show this result for one channel, where one can see the asymptotes corresponding to the energies in the free case. As a consequence, an infinite number of poles are predicted for a particular size of the box. Furthermore, for one channel, we can write the amplitude in infinite volume $T$ for the energy levels $(E_i)$ as 
\begin{equation}
T=(\tilde{G}(E_i)-G(E_i))^{-1}.
\label{eq:bs2}
\end{equation}
These energies have a dependence on $L$ as shown in Fig.~\ref{fig:leddst}, where the energies are determined for the two first levels corresponding to the channel $D^{+}D^{*-}$. 
In Fig.~\ref{fig:leddst}, the two first free energy levels are also shown with dotted lines.

\section{Two channel case}
\begin{figure}
\includegraphics[scale=0.7]{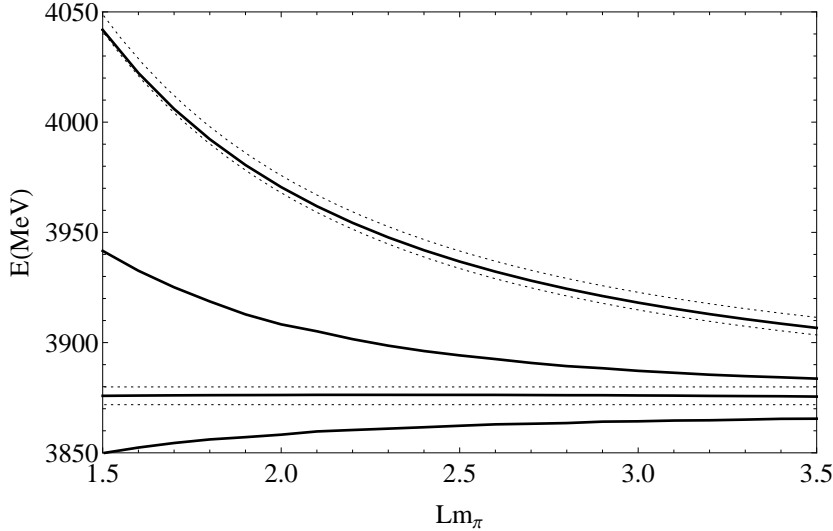}
\caption{$L$ dependence of the energies of the poles for the two first levels of $D^{+}D^{*-}$ and $D^{0}\bar{D}^{*0}$. Dotted lines correspond to the free energies.}
\label{fig:2ch}
\end{figure}

In the previous section we have shown the results for the single channel, $D^{+}D^{*-}$, scattering in a finite box. Next step is to include the $D^{0}\bar{D}^{*0}$ channel. In the work~\cite{Gamermann:2007fi} (see also~\cite{Gamermann:2009fv,Aceti:2012cb}), a pole at $\sqrt{s}=3871.6 $ MeV is obtained using a subtraction constant of $\alpha_H=-1.265$, with a binding energy of 0.2 MeV with respect to the neutral channel When we address the inverse problem in the next section, for the sake of simplicity, we take only two channels, $D^{+}D^{*-}$ and $D^{0}\bar{D}^{*0}$, reevaluating the coupled channel calculation explained in section II (see Table~\ref{tab:x3872}). Then, a new value $\alpha_H=-1.153$ is needed in order to get the same position of the pole. The novelty of this study is the inclusion of two channel in the finite box, where the energies are found using the condition of Eq. (\ref{eq:tpole}). As one can see in Fig.~\ref{fig:2ch} now we have two curves for each level, when for a single channel we had only a trajectory of the energy for each level.
This feature is understood looking into Fig.~\ref{fig:2ch}, where the free energies for the channels $D^{+}D^{*-}$ and $D^{0}\bar{D}^{*0}$ (dotted lines) correspond to the position of the asymptotic lines of Fig.~\ref{fig:vg} for each $L$.
Now new asymptotes appear, with respect Fig.~\ref{fig:vg}, corresponding to the free energies of the $D^{0}\bar{D}^{*0}$ channel. Since the determinant of Eq.~\ref{eq:tpole} has a zero between two asymptotes, the number of bound states in the box is now doubled.
It is interesting to note, by looking at Figs.~\ref{fig:leddst} and \ref{fig:2ch}, that the nondiagonal transition potential between the $D^{+}D^{*-}$ and $D^{0}\bar{D}^{*0}$ has a repulsive effect among the levels, which are now more separated than in Fig.~\ref{fig:leddst}.

\section{The inverse problem}
Once we have determined the dependence of poles of $\tilde{T}$ with $L$ using the potential for the $D D^*$, we want now to study the inverse problem. The idea is that QCD lattice data can be used to determine bound states of the $D \bar{D}^*$ system.
For this purpose we assume that the lattice data are some discrete points on the energy trajectories obtained by us. Starting with a set of synthetic data of energy and $L$, we wish to determine the potential which generates them. Thus, simulating Lattice data, we evaluate the potential, and furthermore, by means of Eq.~\ref{eq:bs} we determine the pole position of the X(3872) in infinite volume with this potential. This study is very useful since we can estimate the uncertainties in the pole depending on the errors of the lattice data. 

\begin{figure}
\includegraphics[scale=0.5]{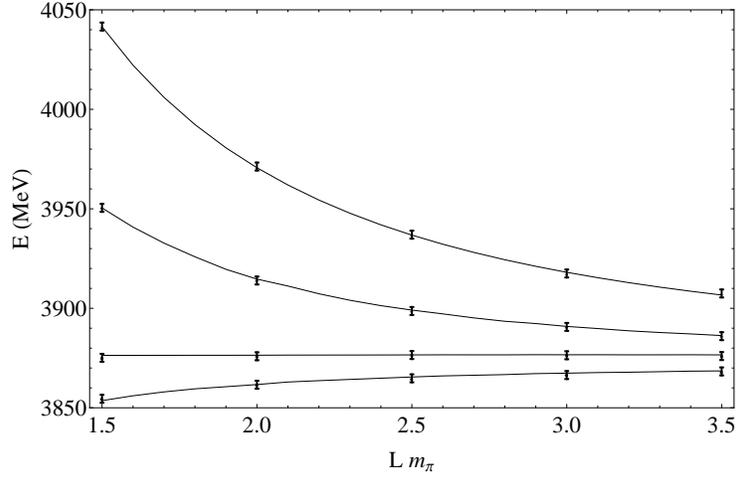}
\caption{Fit to the data. Dots with error bar are the synthetic data generated as explained in the text. Solid lines show the results obtained using the potential fitted to the synthetic data.}
\label{fig:fit}
\end{figure}

Thus, we generate a set of data for some $L$ for a value of the subtraction constant $\alpha=-1.153$. In this case we generate 5 points in a range of $L m_{\pi}=\left[1.5,~3.5\right]$ and take 4 levels, this corresponds to $n$=0 and 1 in the momentum for both channels $D^{+}D^{*-}$ and $D^{0}\bar{D}^{*0}$. In addition, we simulate uncertainties in the obtained data, moving randomly by 1 MeV the centroid of the energies, then we assign an error of 2 MeV to these data. In Fig.~\ref{fig:fit} we show the simulated set of data.

The second step is to choose the potential. We have chosen a potential with linear dependence in $\sqrt{s}$. This is given by
\begin{equation}
V_i=a_i + b_i \left(\sqrt{s}-\sqrt{s^{th}} \right)
\label{eq:potfit}
\end{equation}
where $\sqrt{s^{th}}=m_{D^0}+m_{\bar{D}^{0*}}$ is the energy of the first threshold, and $i$=1, 2 and 3 are the indices for each channel ($i$=1 for $D^{+}D^{*-}$, $i$=2 for $D^{0}\bar{D}^{*0}$ and $i$=3 for the nondiagonal potential). Therefore, there are six parameters to determine in the potential. With all these ingredients, we do the fit, evaluating those values of the parameters in Eq. (\ref{eq:potfit}) that minimize the $\chi^2$ function. In Fig.~\ref{fig:fit} we show the result of this fit together with the error band, which is obtained in the standard method~\cite{Doring:2011vk} varying randomly the parameters of the potential in a moderate range (10$\%$ change) and choosing the set of parameters that satisfy the condition $\chi^2 \leq \chi^2_{min}+1$. With these sets of parameters we determine the binding energy of the system with its dispersion from the pole of $T=(V^{-1}-G^{DR})^{-1}$.
In both $\tilde{T}$ and $T$ we need a value of $\alpha$ to determinate $\tilde{G}$ or $G^{DR}$. The interesting thing that we observe is that the results for the binding energy are essentially independent of the choice of $\alpha$. Changes in $\alpha$ revert on changes of $V$ that compensate for it. We made choices of $\alpha_H$ between -1.2 and -2.2.

\begin{figure}
\includegraphics[scale=0.5]{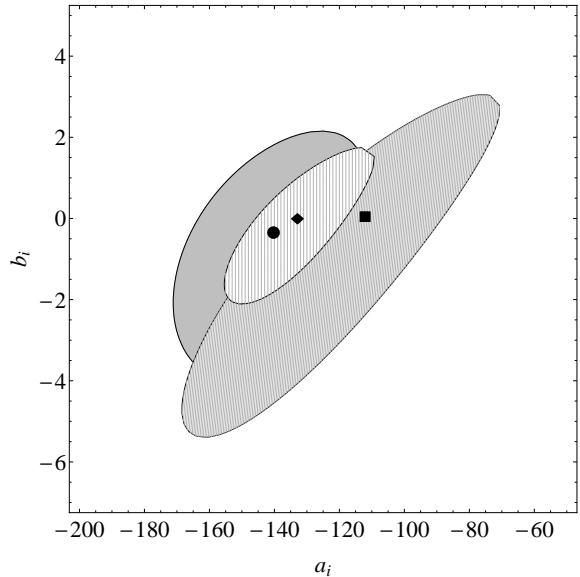}
\caption{Contour plot for the $\chi^2$ representing $\chi^2 \leq \chi^2_{min}+1$. Each area correspond to a pair of parameters $\left\lbrace a_i,b_i\right\rbrace$ for the same potential. Points correspond to values of the parameters in the $\chi^2$ minimum. (Circle and grey area are for $a_1$ and $b_1$, Square and diagonal lined area are for $a_2$ and $b_2$ and Diamond and vertical lined area are for $a_3$ and $b_3$.)}
\label{fig:chi2}
\end{figure}
\section{Results}
In the previous section we have commented our aim to determine the binding energy of the system with its uncertainty depending on the set of data chosen in the analysis. We choose several sets of data from Fig.~\ref{fig:fit}, varying also their assumed errors, and show the results obtained for the energy of the bound state in Table~\ref{tab:respoles}.
We have fitted the first two levels ($n$=0 and 1) for both channels $D^{+}D^{*-}$ and $D^{0}\bar{D}^{*0}$ which gives four branches (B) in the data. The first option is taking only the first level with $n$=0, so we have only two branches, which is more realistic for Lattice results. With this choice we also consider several options of the number of points (P) on $Lm_{\pi}$.
Then we take 5 points ($Lm_{\pi}$=1.5, 2.0, 2.5, 3.0 and 3.5) in one case and 3 points in another one ($Lm_{\pi}$=1.5, 2.5 and 3.5). The last option that we consider is a modification of the error bars of the energies ($\Delta E$) and the variation in the position on the centroid ($\Delta C$). We choose a first set of high precision with $\Delta E$=2 MeV and $\Delta C$=1 MeV and a second, less accurate, set with $\Delta E$=5 MeV and $\Delta E$=2 MeV. We have done the fits for different possible combinations of these variations in the data set up.
\begin{table}
\begin{tabular}{cccc|cccccc|cccc}
\hline
\hline
\multicolumn{4}{c|}{Data}&\multicolumn{6}{c|}{Parameters}&\multicolumn{4}{c}{Results}\\
B&P&$\Delta E$&$\Delta C$&$a_1$&$a_2$&$a_3$&$b_1$&$b_2$&$b_3$&$\chi^2$&Pole&Mean Pole&$\sigma$\\
\hline	
4&5&2&1&-140.18&-112.08&-132.81&-0.310&~0.074&~0.012&2.32&3871.51&3871.49&0.07\\
4&5&5&2&-140.18&-112.08&-132.81&-0.310&~0.074&~0.012&0.79&3871.51&3871.25&0.38\\
4&3&2&1&-133.01&-131.92&-124.60&-0.242&~0.048&-0.075&1.02&3871.44&3871.49&0.18\\
4&3&5&2&-120.09&~-98.19&-150.94&-0.377&-0.075&~0.102&0.28&3871.41&3871.15&0.49\\
\hline
2&5&2&1&-176.08&-154.11&~-89.26&~9.92&~7.01&~-8.72&0.259&3871.70&3871.47&0.30\\
2&5&5&2&-158.49&-152.15&-103.23&~4.56&~6.58&~-6.74&0.982&3871.34&3871.30&0.43\\
2&3&2&1&-132.74&-176.62&-105.53&~3.23&~0.84&~-3.36&0.074&3870.51&3870.48&0.61\\
2&3&5&2&-226.57&-194.51&~-32.74&31.81&13.28&-18.89&0.942&3869.49&3870.37&1.06\\
\hline
\hline
\end{tabular}
\caption{All possible set up changing number of branches ($B$), number of points ($P$), energy error bar ($\Delta E$) and centroid of the energies ($\Delta C$) and their set of parameters fitted. The columns denoted as Results are the $\chi^2$ obtained in the fit, the pole is determined with the parameters, and the mean pole and the dispersion are calculated as explained in the text. The results are for $\alpha=-1.25$. As noted in the text, the use of different values of $\alpha$ change the potential but not the binding energy. Note that we quote values of total $\chi^2$ not the reduced one, which is always much smaller than 1.}
\label{tab:respoles}
\end{table}
The results of the fits are shown in Table~\ref{tab:respoles}, where the first four columns determine the chosen set up of the synthetic data. The next columns are the fitted parameters, value of $\chi^2$ and pole position. The energy values in the ``Pole'' column correspond to the pole positions of the $T$ matrix using the $G^{DR}$ loop function of Eq. (\ref{eq:gpropdr}) together with the parametrized potential of Eq. (\ref{eq:potfit}). To test the stability of the pole with the parameters, we vary randomly the parameters by $10\%$. If the new $\chi^2$ calculated with those parameters is less than the $\chi^2$ obtained in the fit plus one, we determine the pole position, otherwise it will be discarded. We iterate several times until we get 20 or 30 values of the pole positions. Then, we calculate the mean value of those pole positions and their dispersion $\sigma$.

The results are in the line with one should expect: fewer branches, fewer points or bigger errors which reverts into a higher dispersion in the binding energy. Since it is difficult for Lattice simulations to calculate higher levels, we have done also the test for the first level of energies for both channels, and in all cases the dispersion of the pole is higher than in the case where two levels are taken into account.
Since the experimental errors in the binding of the $X(3872)$ are of the order of 0.20 MeV, the exercise done is telling the level of precision demanded for the Lattice data if the experimental precision is to be matched.

\section{Conclusions}

We have studied the $X(3872)$ state using coupled channels $D^{+}D^{*-}$ and $D^{0}\bar{D}^{*0}$ in a finite box. This is done for a small binding energy. In the direct problem, we have reproduced the energy dependence with the size of the box $L$ in the two channel case. We obtain two energy curves for each level corresponding to the neutral and charged channels. On the other hand, we have addressed the inverse problem, obtaining the potential from the simulated lattice data with the aim of using it to evaluate the pole position in the infinite box case. The fit of the different setups give us an idea of what one should expect when analysing Lattice data. First one needs that the fit should be good enough, that is, a chi square function should be sufficiently small. In order to reproduce the small binding energies. In addition, we have observed that in order to get a good precision in the binding energy, one does not need to extract the lattice data with very small errors. Indeed, even with errors in the data points of 5 MeV, one can obtain the binding energy with 1 MeV (or even smaller value) precision. 
However, by looking at rows two and three of Table \ref{tab:respoles} it also becomes clear that very high precision in the binding energy requires small errors in the Lattice data. As seen in Fig.~\ref{fig:fit}, this is necessary to distinguish between the levels of $D^{+}D^{*-}$ and $D^{0}\bar{D}^{*0}$ at large $L$.
From a practical point of view, knowing that it is difficult to get four levels in actual Lattice calculations, it is rewarding to see that with only two levels one can get quite an accurate value for the binding, provided the levels are evaluated at several values of $L$ with enough precision.
We hope that this work gives a reference in the study of Lattice QCD for best strategies in order to obtain optimum values of the binding of the $X(3872)$ state.

\end{document}